\documentclass[aps,prl,twocolumn,groupedaddress]{revtex4}

\usepackage{amsmath}
\usepackage{amssymb}
\usepackage{graphicx}
\usepackage{soul}
\usepackage{siunitx}

\begin{document}

\title{Getting into shape: how do rod-like bacteria control their geometry? }
\author{Ariel Amir}
\email[]{arielamir@physics.harvard.edu}
\affiliation{Department of Physics, Harvard University, Cambridge, MA 02138, USA}
\author{Sven van Teeffelen}
\email[]{sven@pasteur.fr}
 \affiliation{Groupe Croissance et Morphog\'en\`ese microbienne, Institut Pasteur, Paris, France}
\date{\today}

\begin{abstract}
Rod-like bacteria maintain their cylindrical shapes with remarkable precision during growth. However, they are also capable to adapt their shapes to external forces and constraints, for example by growing into narrow or curved confinements. Despite being one of the simplest morphologies, we are still far from a full understanding of how shape is robustly regulated, and how bacteria obtain their near-perfect cylindrical shapes with excellent precision.  However, recent experimental and theoretical findings suggest that cell-wall geometry and mechanical stress play important roles in regulating cell shape in rod-like bacteria. We review our current understanding of the cell wall architecture and the growth dynamics, and discuss possible candidates for regulatory cues of shape regulation in the absence or presence of external constraints. Finally, we suggest further future experimental and theoretical directions, which may help to shed light on this fundamental problem.
\end{abstract}
\maketitle

\section{Introduction}
\label{intro}

\begin{figure}[h]
\includegraphics[width=0.8 \linewidth]{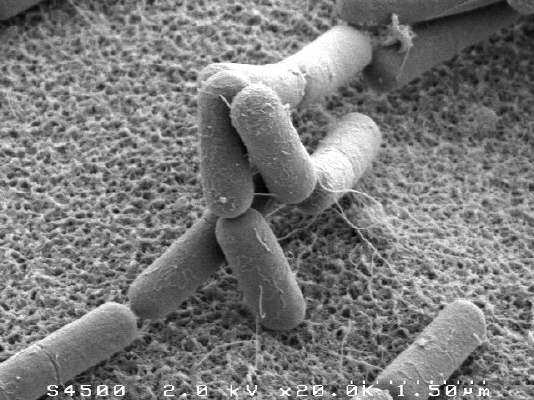}
\caption{Electron microscopy images of \emph{B. subtilis}, taken by Thierry Meylheuc. In a given growth medium, the different bacterial cells have a smooth, highly reproducible cylindrical shape, with relatively small fluctuations in length and radius. The image is reproduced from Ref. \cite{chastanet}, courtesy of A. Chastanet.}
\label{bsubtilis}
\end{figure}

Cells of all organisms and kingdoms face a common challenge of regulating their own shapes to facilitate viability and growth, but also being able to react to external spatial constraints and mechanical forces that eventually require adaptive changes in cell-shape or cellular growth, see \cite{dumais} for an excellent review of the diverse strategies used by organisms with cell walls.  In single-celled bacteria, cell shape is often very precisely controlled, as illustrated in Fig. \ref{bsubtilis}. Bacteria come in a broad range of shapes and sizes (see Fig. 1 of \cite{young2} for a striking graphical representation),  yet despite decades of research, our understanding of how these shapes are controlled and regulated at a molecular level is far from complete. Given the large difference in length scales between the macroscopic cell shape (\si{\um}) and the microscopic proteins, enzymes, and molecules responsible for cell shape (\si{\nm}) -- how is such precise control over shape achieved?

In a given growth medium, various rod-shaped bacteria such as the canonical Gram-negative \emph{Escherichia coli} or the well studied Gram-positive \emph{Bacillus subtilis} elongate while maintaining a constant diameter.  Strikingly, many rod-like bacteria elongate by expanding their cell envelope all along the cell envelope, as compared to growing from the tip only.  These cells maintain their diameter even if cell division is inhibited and cell length reaches dozens of microns \cite{wang}.

%
Here, we focus on this example of rod-like growth, and discuss our current understanding of cell-shape regulation.  The discussion will mainly consist of a physicist point-of-view, where the molecular machinery of cell-wall insertion has been ``coarse-grained''.  We will not discuss in detail the particular action of specific enzymes or the biochemical properties of the peptidoglycan (PG).  Rather, we shall focus on the mechanics of the cell wall and the sensory cues, which might enable the tight regulation of shape.  One important cellular component for shape regulation we will highlight is the bacterial cytoskeleton.

Bacterial cell shape is not only under auto-regulation but is also subject to external mechanical perturbations (such as geometric confinement or external forces).  Cells are known to adapt their growing shapes to these forces \cite{whitesides,dekker}.  Learning about the cellular response to external forces may be important to understand the intra-cellular regulation of shape in unconstrained environments.  The study of auto-regulation and external perturbation of cell shape thus requires an interdisciplinary effort of biologists, physicists and materials scientists, as it requires an understanding of the non-trivial mechanical problems associated with thin, elastic media: while in many cases in biology a qualitative understanding of a phenomenon is sufficient to understand the crux-of-the-matter, shape regulation may involve the sensing of geometric cues and of mechanical stresses and strains, which, in turn, are integral parts of regulatory feedbacks.

There are additional, fascinating questions associated with the intersection of mechanics and bacterial growth, that we shall not discuss here, such as the forces exerted by the Z-ring in the bacterial division process \cite{egan2013physiology,li2013ftsz,piro20133d,sun}, the role of crescentin in shaping curved cells \cite{cabeen}, and the growth of curved and helical bacteria \cite{sycuro2010peptidoglycan,typas2011regulation}, to name but a few. 
\section{Necessity for regulation}
\label{sec:1}
Even for the seemingly straightforward mode of elongation of rod-shaped cells, maintaining the rod shape is a non-trivial task -- simple``templating'' mechanisms where glycan strands are placed in parallel to the existing ones would not be stable to the random fluctuations of growth, especially in light of the disorder in the mesh, which we shall elaborate on further down. To gain intuition, consider a different regulatory problem -- how does a growing leaf stay flat? It turns out that it is a non-trivial task to be flat. It was shown that a negative feedback regulatory circuit is required to avoid a bumpy leaf structure, which is distinctively different from the smooth, flat leaves we are used to and take for granted \cite{sharon}.

For the cylindrical growth, previous works started to tackle this problem by comparing the robustness of various growth mechanisms \cite{wingreen2}, and found that uniformly distributed, helical insertions are quite robust. Yet in their study no strategy was proven to be robust in the true sense, i.e., were we to start from a spherical cell, it is unlikely that the cell would adapt to its rod-shape when using any of the proposed strategies, nor would a rod-shaped cell maintain its diameter over many rounds of division. Bacterial cells do precisely that -- as was shown for \emph{E. coli}: after the cells were significantly distorted when grown in a chamber thinner than their diameter, they recovered their native shape after several rounds of division \cite{dekker}.
A recent study by Ursell \emph{et al.} suggests that cytoskeletal MreB in \emph{E. coli} could play an important part in regulating cell shape. They found that MreB filaments serve as a local sensor of bacterial envelope curvature and thus direct cell-wall insertion to these sites. The authors show that this mechanism could help maintain a cylindrical cell straight \cite{Ursell2014}. Potentially, this same curvature-sensing mechanism could also play an important role in maintaining cell diameter. We will come back to this further down.
In the following we summarize what is known about cell-wall synthesis, the major stress-bearing component of the cell envelope, and how it might lead to stable cell shape.  We will then discuss how auto-regulation and cell-shape response to external forces could come about.

\section{Microscopic cell-wall structure and molecular mode of cell-wall growth}
\label{sec:2}
\begin{figure}
\includegraphics[width=0.8 \linewidth]{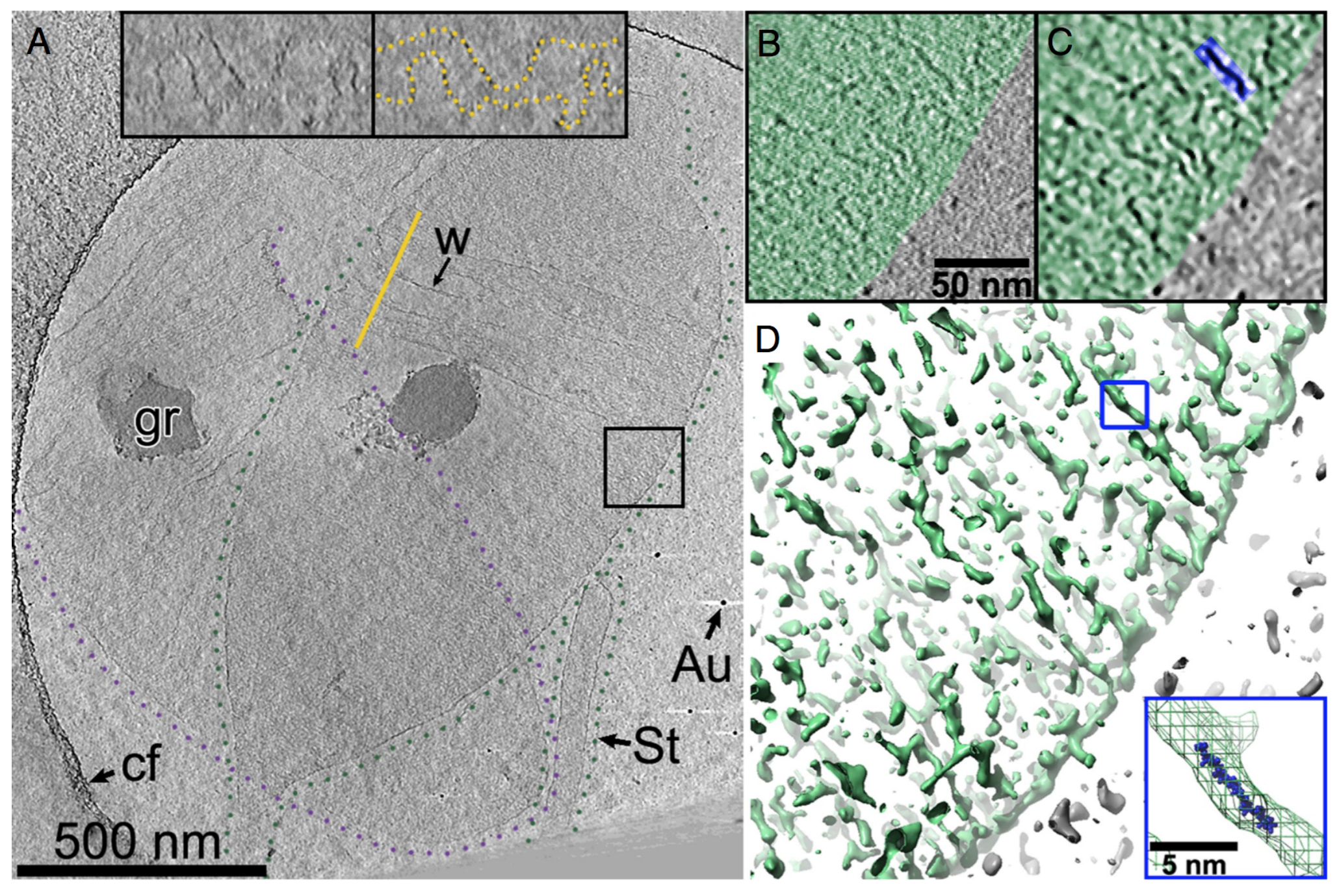}
\caption{Details of peptidoglycan organization as obtained by cryo-electron tomography; reproduced from Gan \emph{et al.} \cite{Gan:2008}; courtsey of Grant J. Jensen. Computational reconstructions of the three-dimensional electron density of a cell-wall sacculus of the Gram-negative bacterium \emph{C. crescentus} reveal a circumferential orientation of the cell-wall glycan strands:  Shown are  two overlapping cell-wall sacculi (A), outlined by green or violet dotted lines. The boxed region in (A) is magnified in panels (B-D). Panel (D) is an iso-density plot, which shows long circumferentially oriented structures that are presumably individual glycan strands. The inset in (D) displays a superposition of the blue-boxed glycan strand and an atomic model of a 9-subunit-long glycan strand, for comparison of scale.}
\label{fig:Gan}
\end{figure}

The bacterial cell shape is physically determined by the PG cell wall, a covalently bonded network of sugar strands cross-linked by short peptide bridges.  The rigid PG meshwork counteracts the high turgor pressure set by the difference in osmotic potentials between the cell and its environment \cite{Deng:2011vu}.  In Gram-negative bacteria cryo-electron tomography images of isolated cell-wall sacculi suggest that the PG forms a monolayer with glycan strands running in a near-circumferential direction around the long axis of the cell \cite{Gan:2008} (see Fig. \ref{fig:Gan}).  This observation is in agreement with atomic force microscopy (AFM) measurements on isolated cell-wall sacculi \cite{yao}, which have revealed that the elastic constants of the cell wall are anisotropic. This anisotropy is expected because of the difference in stiffness between rather rigid circumferentially oriented glycan strands and the comparatively floppy peptide bonds. Interestingly, there is also a two-fold difference of cell-wall mechanical stresses between the circumferential and axial directions that comes about due to the cylindrical geometry of the cell. Theoretical modelling suggests that the large turgor pressure drives the cell wall elasticity to the non-linear regime \cite{boulbitch,shaevitz2}.

In Gram-positive bacteria the cell wall is much thicker than in Gram-negatives (e.g., in \emph{B. subtilis} the cell wall is approximately 30 \si{\nm} thick \cite{Beeby:2013bf,Misra:2013ie}).  The thickness of the cell wall has prevented molecular-resolution imaging of the glycan strands. However, recent cryo-electron microscopy and surface atomic force microscopy experiments have revealed circumferential furrows in the cell-wall surface \cite{Beeby:2013bf,Andre:2010bx,Hayhurst:2008ei} with a spacing of roughly 50 \si{\nm}.  While this observation is in agreement with the model of circumferential glycan strands it also suggests a higher-order three-dimensional structure, which is not understood yet.  For both Gram-positive and Gram-negative bacteria biological text books often depict the architecture of the PG cell wall as a regular lattice. However, the short length of the glycan strands of several \si{\nm} \cite{harz} suggests that the structure is much more disordered.

The structure of the newly synthesized cell wall has been revealed in a different, indirect manner:  Proper cell-wall synthesis depends on the bacterial cytoskeleton, particularly on one or multiple isoforms of the widely conserved actin-homologue MreB \cite{shaevitz2010structure}. MreB forms filaments in the cytoplasm that are attached to the cytoplasmic membrane in both Gram-negative \cite{Kruse:2003kz} and Gram-positive bacteria \cite{Jones:2001wi,Olshausen:2013cw,Reimold:2013ea}.  The length of the MreB filaments is currently under debate, in particular because native-expression-level MreB filaments have not been detected in whole cells by electron microscopy \cite{swulius2012helical}.  Irrespective of their exact length, it has been shown by fluorescence microscopy that MreB filaments rotate around the long cell axis in a processive manner in Gram-negative \cite{shaevitz} and Gram-positive bacteria \cite{garner,escobar,Olshausen:2013cw,Reimold:2013ea}.  This rotation depends on PG synthesis and proceeds at a speed compatible with processive insertion of single glycan strands into the PG meshwork \cite{shaevitz}, as already suggested by Burmann and Park in the 1980s \cite{burman}.  It is thus plausible that MreB filaments are physically linked to the enzymes responsible for cell-wall insertion.  In fact, some of the cell-wall-synthesis enzymes have been seen to move in a similar manner as MreB filaments in the Gram-positive \emph{B. subtilis} \cite{garner,escobar}, supporting the hypothesis of physical interaction.
In Gram-negative \emph{E. coli}, at least one important synthesising enzyme, the transpeptidase PBP2, moves rapidly and diffusively, showing no processivity on the sub-second time scale \cite{Lee2014}, thus suggesting a more transient interaction of the cell-wall synthesis proteins.
Ref. \cite{tuson} finds that the timescales, at which disrupting MreB affects cell wall elasticity are similar to the growth time, in consistence with this interpretation.  Furthermore, filaments have recently been reported to move with a filament-length dependent speed \cite{Olshausen:2013cw}.  The speed-length relationship observed is compatible with a simple model of synthesis complexes effectively consituting motor proteins that randomly attach to MreB filaments and exert a force in either of the two circumferential directions.  Accordingly, the speed as a function of length displays a maximum at finite filament length of a few hundered nanometers \cite{Olshausen:2013cw}.

Interestingly, the trajectories of cytoskeletal filaments observed in \emph{E. coli} are slightly helical on average \cite{shaevitz}, suggesting an average helical organization of the cell wall as a whole. This helicity of the cell wall has since been supported by experiments of combined microscopy and optical trapping \cite{Wang:2012fx}:  Wang \emph{et al.} attached fluorescent beads to the envelope of elongated \emph{E. coli} or \emph{B. subtilis} cells using optical tweezers.  They then tracked the bead position before and after osmotically up-shocking the cells in a flow cell.  First, they find that the cells shrink much more along the long axis than along the radial direction -- in accordance with the aforementioned anisotropy of elastic constants.  Furthermore, they also find that the beads follow helical trajectories during the shape transition, which suggests a slight helical anisotropy and, thus, a helical orientation of the PG meshwork -- in agreement with the helical trajectories of MreB motion.  With combined fluorescence microscopy on MreB filaments and computational elastic-network simulations (coarse-grained molecular-dynamics simulations), the authors argue that the helicity might be caused by the orientation of MreB filaments below the cylindrical surface of the cell wall.  The orientation of the MreB filaments with respect to the cell envelope, in turn, could be caused by the filament-intrinsic curvature and twist in combination with a curved surface of the cylindrical cell envelope\cite{Wang:2013fl,andrews2007mechanical}.

Linking cell-wall synthesis to MreB filaments is very interesting from a physics perspective: Multiple independent studies have suggested that MreB filaments assume on average macroscopic lengths of few hundred nanometers \cite{Olshausen:2013cw,Reimold:2013ea,Kruse:2004dx,Jones:2001wi}.  The mechanical stiffness of these filaments \cite{wang} could facilitate the macroscopic organization of the cell-wall synthesis machinery and might thus provide a key ingredient for a robust cell-shape feedback mechanism (see discussion below).  Computational simulations by Furchtgott \emph{et al.} have already shown that the stiffness of MreB could provide the cell with a mechanism to avoid an unfavorable positive feedback of macroscopic cell-wall bulges, i.e., local departures from the intended perfectly cylindrical geometry \cite{wingreen2}.  Their argument goes as follows: If cell-wall insertion was only dependent on the availability of PG substrate, i.e., if new PG was inserted with equal probability at any potential site of insertion, local cell-wall bulges would grow, as they contain a higher number of potential insertion sites.  Conversely, sites with lower cell-wall density would be depleted of cell-wall material, while the surrounding meshwork would expand.  The cytoskeleton could render insertion independent of the local density of PG, simply by bridging small deviations from the cylindrical envelope due to polymer stiffness. Related ideas regarding the role of MreB in stabilizing cylindrical growth are provided in Refs. \cite{sun_stability,sun}, which also illustrate theoretically and experimentally a mechanical instability which can occur in the absence of MreB.
However, while this mechanism could prevent local deviations from a flat cylinder surface regulation of cell shape requires a mechanism that measures large-scale deformations of the cell envelope -- either directly in form of cell-envelope curvature (suggested by Ursell \emph{et al.} \cite{Ursell2014}) or indirectly, e.g., via a modified mechanical stress in the cell wall (see discussion below).

\section{Understanding cell-shape regulation by cell-shape perturbations}
\label{sec:3}
\begin{figure}[h]
\includegraphics[width=0.8 \linewidth]{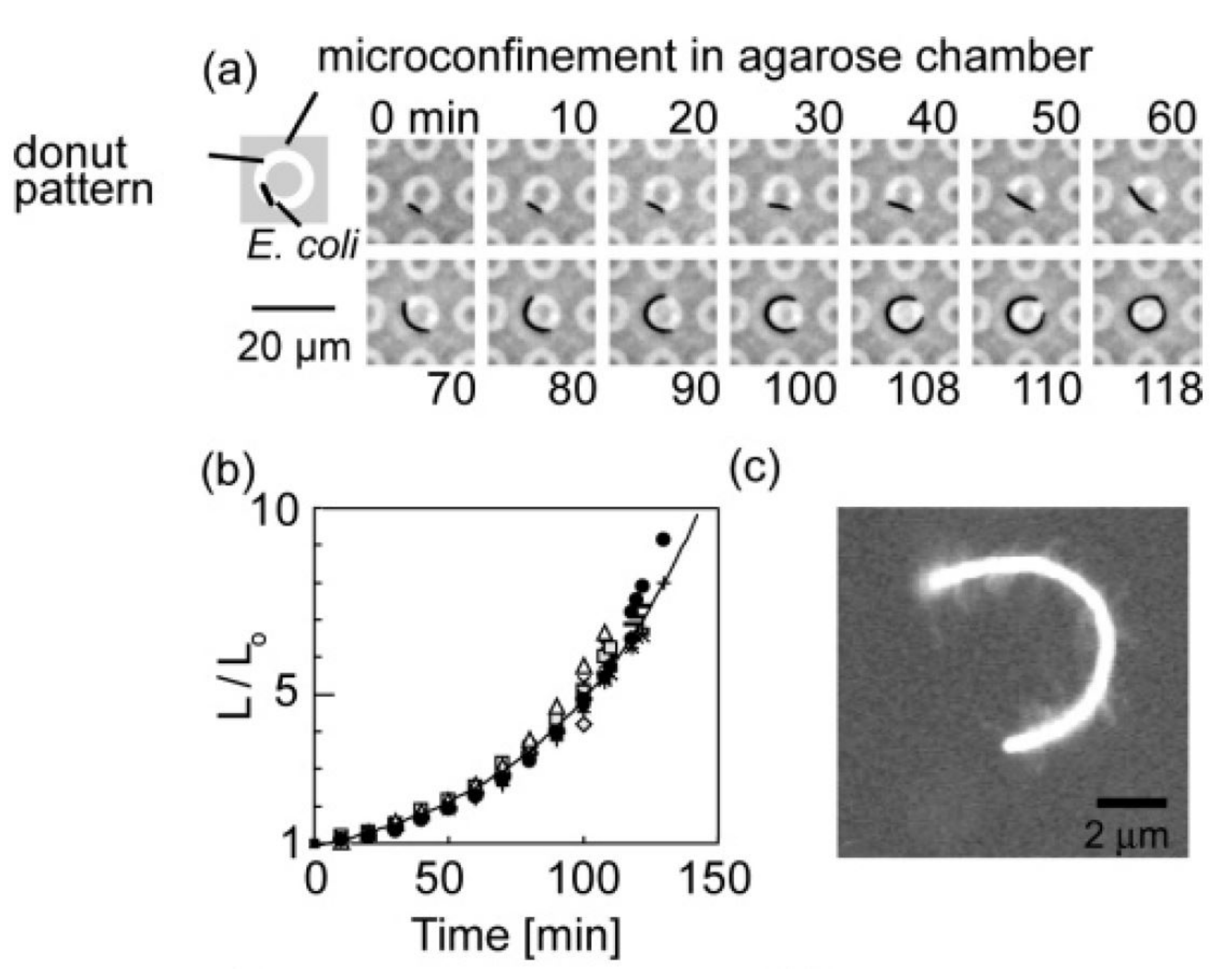}
\caption{Microscopy image of \emph{E. coli}, reproduced from Ref. \cite{whitesides}, courtsey of G. M. Whitesides. (a) The cell was grown in a narrow circular channel. (b) The cell length grows exponentially in time with the physiological growth rate, showing that the cell is at least locally close to its normal growth conditions, yet the cell adapts to the shape of the channel. (c) When taken out of the channel the cell maintained a deformed shape, illustrating that the cell is able to adapt its shape to the confinement during growth, without building up large one-sided stress that would relax by straightening after release from the chamber.}
\label{whitesides}
\end{figure}

Looking at sub-cellular components such as the PG cell wall and the MreB cytoskeleton have fundamentally improved our understanding of the organizing principles of the cell wall in the steady state of rod-like growth.  A different approach to understanding cell-shape regulation is to perturb cylindrical cell shape and observe how the cell reacts to the perturbation -- both during and after the perturbation \cite{bending,whitesides,dekker,jacobs}.  Such an approach is particularly appealing from a physics perspective, as the cell wall is a partially ordered elastic sheet that undergoes elastic and plastic deformations upon mechanical forces and during growth, respectively \cite{bending}. We note that by ``plastic" we mean irreversible due to a change of the covalent peptide and glycan bonds. This change comes about due to the cleaving of existing bonds through enzymes (as opposed to ripping) and possibly through the insertion of new PG material.  Thanks to the non-uniform, possibly adaptive growth process the residual stresses in the cell wall can be much smaller than in plastic deformations happening in non-living materials (e.g: metals).  This allows for the controlled test of molecular and physical models of cell-wall insertion and cell-wall elastic properties.  Besides helping us to understand cell-shape auto-regulation during normal growth, cell-shape deformation experiments also allow us to study how the cell reacts to mechanical and geometric constraints, such as confining spaces (see Fig. \ref{whitesides}).  Ultimately, the molecular mechanisms underlying cellular response to perturbations and the mechanisms underlying cell-shape auto-regulation might be the same; however there might also be components of adaptation and auto-regulation, respectively, that compete against each other, a possibility that we shall later elaborate on.

In a first such experiment, Takeuchi \emph{et al.} found that filamentous \emph{E. coli} cells grown in small, cylindrical confining chambers maintain their shapes after release from the chamber (see Fig. \ref{whitesides}).  Thus, \emph{E. coli} is able to adapt its shape instead of growing as a rod-like cylindrical cell when grown in confined environments.  In another experiment M\"annik \emph{et al.} observed that \emph{E. coli} cells can grow and even divide in shallow confining slits. In both cases cells revert their shapes after sufficient additional growth outside the confining geometry \cite{dekker}.

For another beautiful example of a biophysical approach, consider the work of Ref. \cite{jacobs}: in this work the naturally curved gram-negative bacterium \emph{Caulobacter crescentus} is manipulated genetically to be straight. The dynamics of the straightening process, when starting from a curved cell, was measured in time using optical microscopy. Careful analysis of the mechanics involved led the authors to rule out several models for the observed straightening, and to conclude that processive, circumferential insertions of glycan strands into the cell wall at random locations explain their measurements -- consistent with the more recent and more direct evidence of circumferential insertion as described in Sec. \ref{sec:2}. The differential geometry and mechanics used in this work is far from the standard toolbox of a biologist, yet the conclusions reached are intuitive and understandable -- as well as highly relevant -- to any biologist interested in bacterial morphology.

\begin{figure}[h]
\includegraphics[width=0.8\linewidth]{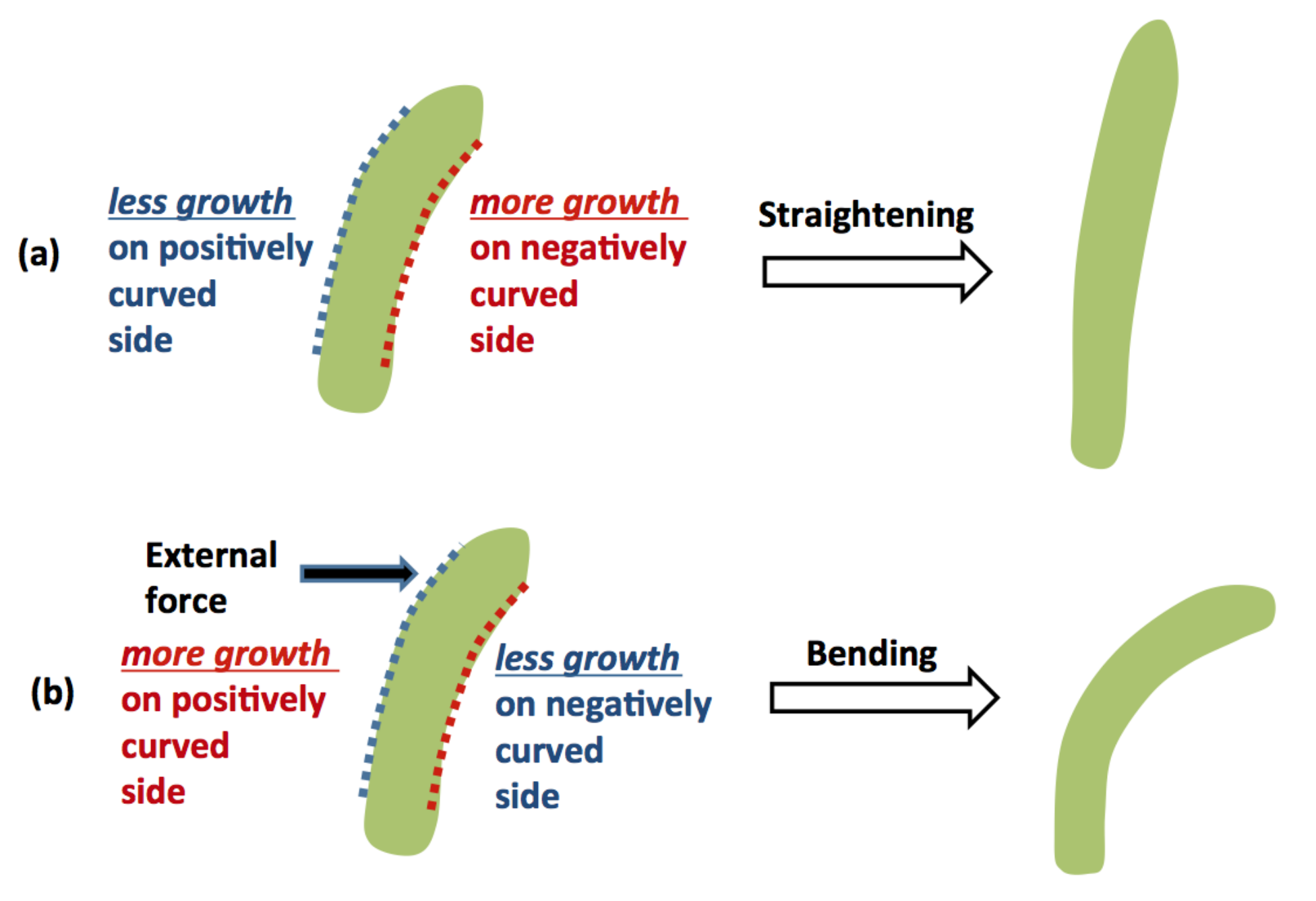}

\caption{Experiments suggest that cell wall growth depends on the situation: typically, cells would grow in a manner that would straighten them, as shown in (a). However, when under external mechanical stress (are a result of a confining environment \cite{whitesides} or an applied bending torque \cite{bending}), the cells would plastically deform to adapt to the new environment. }
\label{schematic}
\end{figure}

In a recent study, a large bending torque was applied to growing filamentous \emph{E. coli} and \emph{B. subtilis} cells \cite{bending}, using a viscous drag, in order to study the elastic and plastic deformations of the cell wall during growth. The authors concluded that mechanical stresses are involved in the regulation of shape in \emph{E. coli}:  the cell grew more cell wall on the side of the flow, where a tensile stress stretched the cell wall, and grew less on the opposite side where the external stress was compressive (see Fig. \ref{schematic}). Thus, the cell reacts to an external force by adapting its shape. The observed plastic shape deformations during growth were consistently interpreted in terms of the dislocation-mediated growth theory \cite{nelson_review,amir_nelson_pnas}. In this formalism the circumferential insertions are interpreted in terms of edge dislocations in the PG mesh, building on concepts developed in the context of the physics of defects in metals.  This mechanism of plastic deformations might also be responsible to the circular cell shapes observed in cells grown in confinement (Fig. \ref{whitesides}), and is reminiscent of the role of ``smart autolysins" proposed by Koch \cite{koch}.

Interestingly, when the external stress (due to the flow) is switched off, the cell straightens.  While this seems in accord with the previously described experiment on \emph{C. crescentus} straightening, there is a crucial difference between the two: in the previous experiment the curvature of the cell centerline was measured to decay exponentially during the straightening process, however, at a rate lower than the rate of exponential cell elongation due to growth.  Thus, the filamentous cells never reached a straight configuration.  According to the authors' model of random processive insertion the decay of the centerline curvature is a monotonically increasing function of the length of newly inserted glycan strands (the amount of processivity).  Yet, even for glycan strands much longer than the cell circumference (infinitely processive insertions) the decay rate would saturate at a finite value (which happens to be the growth rate).  The shape of the bacteria (were it not to divide) would be self-similar \cite{mukhopadhyay}, i.e., a curved cell, which does not divide would maintain a curved shape.  On the contrary, filamentous \emph{E. coli} cells in the flow-cell experiment straighten more than by the maximum straightening rate in the random-insertion process.  This result suggests the presence of an additional straightening mechanism. Further work is needed in order to establish whether this is an ``active" mechanism, through which \emph{E. coli} attempts to ``correct" for cell shape deformation, or a result of the coupling to the residual mechanical stresses in the cell wall, which persist even after the external force is switched off. The emerging picture of cell bending and straightening in \emph{E. coli} is illustrated in Fig. \ref{schematic}, showing the way a cell responds to curvature and external forces.

\section{Possible feedback mechanisms}
\label{sec:5}	
All experiments and computational simulations described above leave us with two questions:  How -- mechanistically -- does the cell adapt its shape to the influence of external forces, and secondly, how does it restore and auto-regulate cylindrical shape during normal growth?

The plastic deformation of the bacterial cell shape during long-term application of a torque suggests that the bacterium preferentially inserts new PG material on the side of the cell facing the flow (where it experiences a higher stress in its cell envelope).  Alternatively, the cell could grow a less dense PG meshwork on the flow-facing side.  In either case the local cell-wall synthesis machinery must react to the mechanical stresses applied.

How do cells maintain their cylindrical shape or re-acquire it after perturbation?  Recent work by Ursell \emph{et al.} \cite{Ursell2014} suggests
that cells are able to sense the local cell-envelope curvature through cytoskeletal MreB filaments. As discussed above, membrane-associated MreB filaments are stiff and could thus favor their own localization at sites of particular cell-envelope curvature, thus effectively constituting a curvature sensor. Ursell \emph{et al.} find that MreB localizes at positions of cell-wall indentations, i.e., at negative Gaussian curvature of the envelope. Monitoring the local expansion of the cell wall (using a cell-wall stain as fiducial marker) and imaging both cell shape and the localization of MreB at the same time, Ursell \emph{et al.} propose that MreB filaments are physically linked to sites of PG insertion.  Thus, MreB may guide the PG insertion machinery to sites of preferred curvature. Furthermore, they found that these regions of the cell envelope flipped curvature sign after PG insertion.  The curvature-based insertion scheme could thus provide a way to help maintain cell shape during rod-like growth by providing an inherent feedback mechanism between the PG-insertion machinery, which determines cell shape, and the cell shape, which, in turn, determines the location of the PG-insertion machinery. Indeed, a coarse-grained computational simulation suggests that coupling processive cell-wall insertion to cell-envelope curvature helps keeping a cell straight \cite{Ursell2014}.

However, the curvature-based growth mechanism alone cannot account for the aforementioned bending experiments \cite{bending}: if only curvature sensing is present, upon being elastically deformed to the right, the cell would attempt to add more material on its right side, since it has a negative curvature. This implies that when the external force which led to the bending is switched off, the cell should be deformed to the left – since more material was inserted on the right hand side. However, the experiments show that the cell is deformed to the right. Therefore the sign of the differential growth expected from a curvature-based mechanism is opposite of what is experimentally observed. Further work is needed in order to establish the connection between these two observations, and the relative importance of mechanical stress and geometric curvature.

%

\section{Future prospects}
\label{sec:4}
The previous examples of cell-shape experiments and modeling illustrate the effectiveness of combining theoretical modelling and novel experimental techniques to improve our understanding of cell wall regulation and the dynamics of growth. There is lots of room to further explore both of these avenues.

On the theoretical side, the attempts to study the robustness of growth have been primarily numerical. Computational simulations from KC Huang's lab have demonstrated how global helical cell-wall structure and local cell-wall integrity can emerge from mesoscopic cytoskeletal filaments \cite{wang}.  However, the computational resources, which are at our disposal at the moment do not allow for the modelling of the full number of interacting units in the peptidoglycan mesh, and can only provide intuition as to the true robustness of a particular model.  Alternative more ``coarse-grained" approaches have been recently introduced \cite{nelson_review,amir_nelson_pnas,amir_nelson_PRE}, in which the relatively small number of active growth sites correspond to moving dislocations in the peptidoglycan mesh, yet in these previous studies a perfectly cylindrical geometry was assumed. In general, the theory describing thin interfaces such as the bacterial cell wall, shallow shell theory, involves highly non-linear partial differential equations, making analytic progress challenging. One possible direction would be to adapt the existing equations, which are commonly used by engineers to study thin shells (the Donnell-Mushtari-Vlasov equation, which generalize the F\"{o}ppl-von K\'{a}rm\'{a}n equations) to incorporate \emph{growth}, and test the stability of the equations to perturbations using linear stability analysis. A second theoretical tool, which was recently introduced is the use of a metric to describe curved surfaces \cite{incompatible2}. The non-uniform growth can be cast in terms of its effect on the ``target metric", and for a thin interface the shape is determined by the Gaussian curvature of that metric. This tool has proven useful in calculating the metric necessary to achieve a desired shape, which can then be prescribed onto a thin polymer sheet, leading to remarkable control of three-dimensional objects \cite{klein,kim}. The deformations of bacterial cells in the microfluidic experiment described above \cite{bending} can, in fact, also be described using the effect on the Gaussian curvature of a metric \cite{jayson}, and this could be a powerful theoretical tool to handle the problem of cylindrical stability.

On the experimental side, it seems that further research is needed in order to establish the relative role of both curvature-related and stress-related regulation. Repeating both  microfluidic experiments described in Sec. \ref{sec:3} while following the dynamics of MreB would provide more information regarding the differential growth. With new possibilities to track the metric of the cell wall directly in live cells \cite{Ursell2014} we can now quantitatively understand where new material is being deposited and correlate it with the stress distribution on the cell wall. Similar approaches in the very different context of tissue morphogenesis have proven useful; remarkably, also in this case, mechanical stress have been shown to play an important regulatory role \cite{tissue}. Making even larger perturbations is another experimental route, which may lead to new insights: both in gram-positive and gram-negative bacteria the production of cell wall can be damaged such that in a low osmolarity medium the cell is still viable in spite of lack of cell wall, leading to spherical cells \cite{deboer}. Recently, recoveries from such spherical cells into the native rod-shaped forms have been observed in \emph{B. subtilis} \cite{ethan}. How does a sphere grow to be a cylinder? The observed path of recovery shows a distinct morphology, which provides important constraints for theoretical models -- not only do they have to predict a robust cylindrical growth, but the form of recovery must also agree with these experimental findings.

 A complete theory of bacterial cell shape should also account for the magnitudes of both radius and length; the regulation of these two is, however, of very different nature: References \cite{size1,size2,amir_size} suggest a robust mechanism of maintaining cell length in bacteria, consistent with the experimentally observed correlations and distributions, invoking a simple biophysical mechanism that does not couple to mechanics or curvature. This mechanism is obviously decoupled from that of radius maintenance, as is proven by the possibility of having extremely long filamentuous cells, which nevertheless maintain their constant radius \cite{bending}. Various approaches have been used to explain the origin of the micron-scale diameter of \emph{E. coli} and \emph{B. subtilis}, including an energy minimization scheme \cite{sun_PRL} and the natural curvature of MreB filaments \cite{Ursell2014}.

In contrast to \emph{E. coli} and \emph{B. subtilis} various bacteria such as Mycobacteria, Streptomyces, are tip-growers \cite{flardh2012regulation}. How is rod-shape maintained for tip-growers? It is plausible that a different mechanism will be necessary in this scenario. In this case drawing an analogy with the growth of plants and fungi could be helpful, since they organisms are also tip-growers \cite{tip1}. Extensive work has been done on modelling tip-growth and the role of mechanics \cite{tip2,tip3}, and it is intriguing to see whether these concepts could apply for bacteria as well.

We are still far from unravelling the fundamental ``engineering" challenges that biology has to overcome in shaping single cells as well as multi-cellular tissues. Yet the rapid development of new theoretical, computational and experimental techniques in these fields, combined with the recent fruitful collaborations between biologists, physicists and engineers, suggest a promising and exciting future.

\begin{acknowledgements}
AA was supported by the Harvard Society of Fellows and the Milton Fund. SvT was supported by a Human Frontier Science Program Postdoctoral Fellowship.
The authors acknowledge useful discussions and feedback regarding the manuscript from  E. Efrati, O. Amster-Choder, Y. Eun, K. C. Huang, D. R. Nelson, J. Paulose and T. Ursell.

\end{acknowledgements}


\end{document}